\documentclass[11pt]{article}
\usepackage{blois,epsfig}

\def\be{\begin{equation}}
\def\ee{\end{equation}}
\def\bea{\begin{eqnarray}}
\def\eea{\end{eqnarray}}

\begin{document}

\vspace*{2cm}
\title{QCD Pomerons at Non-Zero Temperature}

\author{H. J. de Vega$^a$ and L. N. Lipatov$^b$}

\address{$^a$ Laboratoire de Physique Th\'eorique et Hautes 
Energies, Universit\'es Paris VI et VII, BP 126,\\
4, place Jussieu, F-75252 Paris cedex 05, France.\\
Laboratoire Associ\'e au CNRS UMR 7589. \\
$^b$ Theoretical Physics Department,
Petersburg Nuclear Physics Institute,\\
Gatchina, 188300, St. Petersburg, Russia}

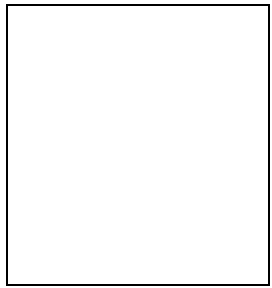
\maketitle\abstracts{We consider the QCD scattering amplitudes at high energies $\sqrt{s}$
and fixed momentum transfers $\sqrt{-t}$ in the leading logarithmic
approximation at a non-zero temperature $T$ in the $t$-channel.  It is
shown that the BFKL Hamiltonian has the property of holomorphic
separability. The holomorphic Hamiltonian for $n$-reggeized gluons at 
temperature $T$ is shown to  be obtained from the $T=0$ Hamiltonian by
an unitary transformation. The connection with field quantization
in accelerated frames and black hole backgrounds is discussed.
The effect of the running of $\alpha_{QCD}$ on the pomeron eigenvalues is investigated.}

\section{Quark Gluon Plasma and Reggeons}

Current theoretical understanding of the quark-gluon plasma (QGP)
generated in heavy nucleus collisions
suggests that the QGP thermalizes via parton-parton
scattering. The QGP is understood to  cool down by hydrodynamic
expansion till the temperature reaches the hadronization scale $\sim 160$MeV.
One interesting phenomena is the suppression of the $\psi $-meson
production in the heavy nucleus collisions due to the
disappearance of the confining potential between $q$ and $\bar{q}$ at
high temperature \cite{QGP}. 
A similar effect should exist for glueballs
constructed from gluons. Because the Pomeron is considered as a composite
state of reggeized gluons, the influence of the temperature on its
properties is of great interest. We constructed in  ref.\cite{PL} the BFKL
equation at temperature $T$ in the center-mass system of the $t$
-channel (where $\sqrt{t}=2\epsilon $) and show that the BFKL dynamics in a 
thermostat for composite states of $n$ reggeized gluons in multi-colour QCD
is integrable. 

Let us consider the Regge kinematics in which the total particle
energy $\sqrt{s}$ is asymptotically large in comparison with the
temperature $T$. In this case one can neglect
the temperature effects in the propagators of the initial and intermediate
particles in the direct channels $s$ and $u$. But the momentum transfer
$|q|$ is considered to be of the order of $T$ (note, that $q_{\mu }$
is the vector orthogonal to the initial momenta $q_{\mu }\approx
q_{\mu }^{\perp }$ ). As it is well known the particle wave
functions $\psi (x_{\mu })$ at temperature $T$ are periodic in the
euclidean time $\tau =i\,t$ with period $1/T$.

We introduce the temperature $T$ in the center of mass frame of the
$t$-channel. Thus, the euclidean energies of the intermediate
gluons in the $t$-channel become quantized as
$$
k_{4}^{(l)}=2\pi l\,T \; .
$$
In the $s$-channel the
invariant $t$ is negative and therefore the analytically continued 4-momenta
of the $t$-channel particles can be considered as euclidean vectors. It
means, that at temperature $T$, the wave functions
for virtual gluons are periodic functions of the 
holomorphic impact-parameter
$\rho =x+iy$ 
with imaginary period $\frac{i}{T}$. Also,
the canonically conjugated momenta $p$ have their imaginary part quantized,
\begin{equation}
\rho \rightarrow \rho +\frac{i}{T} \quad , \quad p=\mbox{Re}\, p+ 
\pi\, i \, l \, T \; .
\end{equation}
with integer $l$ (note that $p=(p_1+ip_2)/2 $).

\section{Reggeon Hamiltonian at temperature $T$}

The calculation of the Regge trajectory $1+\omega (t)$ of the gluon 
at temperature $T$ in the t-channel,
in one-loop approximation reduces to the integration over the
real part $k_{1}$ of the transverse momentum $ k $
of the virtual gluon and to the summation over its imaginary part
$k_{2}=l$. In such a way we obtain
the following result for the trajectory having the separability
property [cf. \cite{lipa}],
\begin{equation}
\omega (-\vec{q}^{2})=-\frac{g^{2}}{8\pi ^{2}} \; N_{c} \; \Omega (-\vec{q}
^{2})\quad , \quad \Omega (-\vec{q}^{2})=\Omega (q)+\Omega (q^{\ast }) \; .
\end{equation}
Here,
\begin{equation}
\Omega (q)=\frac{\pi T}{\lambda } + \frac{1}{2}\,\left[ \psi (1+\frac{iq}{2\pi\,T})
+\psi(1-\frac{iq}{2\pi\,T})-2\psi (1)\right] \; , \nonumber
\end{equation}
where we regularized the infrared divergence for the zero mode $l=0$
introducing a mass $\lambda $ for the gluon (see \cite{lipa}).

A similar  divergence appears in the Fourier transformation $G(
\overrightarrow{\rho }_{12})$ of the effective gluon propagator $(\vec{k}
_{\perp }^{2}+\lambda ^{2})^{-1}$ contained in the product of the effective
vertices $q_{1}k^{-1}q_{2}^{\ast }$ for the production of a gluon
with momentum $k_{\mu }$ (cf. \cite{lipa})
\begin{equation}
G(\overrightarrow{\rho }_{12})=-\frac{\pi \,T}{\lambda }+\ln \left( 2\sinh
\pi \, T \, \rho _{12}\right) +\ln \left( 2\sinh \pi \, T \, \rho _{12}^{\ast }
\right) \; .
\end{equation}
Therefore, the divergence at $\lambda \rightarrow 0$ cancels in the
sum of kinetic and potential contributions to the BFKL equation and the
Hamiltonian $H_{12}$ for the Pomeron in a thermostat has the property of
holomorphic separability with the holomorphic Hamiltonian given below 
(as it is the case for zero temperature \cite{lipa})
\begin{equation}
h_{12}=\sum_{r=1}^{2}\left[ \frac12 \; \psi (1+\frac{ip_{r}}{2\pi\,T})+
\frac12 \; \psi(1-\frac{ip_{r}}{2\pi\,T}) + %\right. \cr \cr &+&\left. 
\frac{1}{p_{r}}\,\ln \left( 2\sinh \pi \, T \,\rho_{12}\right) \,p_{r}-
\psi(1)\right] \; .  \label{hamilT}
\end{equation}
We constructed in ref. \cite{PL} the eigenfunctions and eigenvalues of
the hamiltonian eq. (\ref{hamilT}) in terms of hypergeometric functions.

\section{Wavefunctions in coordinate space}

The Pomeron wave function can be constructed directly in coordinate
space. For this purpose we use the conformal transformation
\begin{equation}\label{exp}
\rho_{r}^{\prime }= \frac1{2 \, \pi \, T} \; e^{2 \, \pi \, T \, \rho _{r} }
\end{equation}
to map the zero temperature wave functions into the wavefunctions at temperature
$T$.

Thus, the Pomeron wave function at non-zero
temperature having the property of single-valuedness and periodicity
takes the form
\begin{equation}
\Psi ^{(m,\widetilde{m})}(\vec{\rho _{1}},\,\vec{\rho _{2}},\,\vec{\rho_{0}})= 
\left( \frac{\sinh [\pi \, T \, \rho_{12}]}{2\sinh [\pi \, T \, \rho _{10}]
\,\sinh [\pi \, T \, \rho _{20}]}\right) ^{m} \times 
\left( \frac{\sinh [\pi \, T \, \rho
_{12}^{\ast }]}{2\sinh [\pi \, T \, \rho _{10}^{\ast }] \,\sinh [\pi \, T \, \rho
_{20}^{\ast }]}\right) ^{\widetilde{m}}\,.
\end{equation}
The orthogonality and completeness relations for these functions can be
easily obtained from the analogous results for $T=0$ (see \cite{lipa})
using the above conformal transformation. 

Moreover, the pair BFKL Hamiltonian $h_{12}$ can be expressed in terms of
the BFKL Hamiltonian at zero temperature in the new variables
\begin{equation}
h_{12}=   \ln (p_{1}^{\prime }\,p_{2}^{\prime })+   %      \\ \cr &&+
\frac{1}{p_{1}^{\prime }}
\,\log (\rho _{12}^{\prime })\,\,p_{1}^{\prime }+\frac{1}{p_{2}^{\prime }}
\,\log (\rho _{12}^{\prime })\,\,p_{2}^{\prime }-2\psi (1)\,, \nonumber
\end{equation}
where $p_{r}^{\prime }=i\frac{\partial }{\partial \rho _{r}^{\prime
}}$. We recognize here the zero temperature BFKL Hamiltonian \cite{lipa}.
In the course of the derivation the 
following operator identity (see \cite{lipa})
\[
\frac{1}{2}\left[ \psi \left(1+z\frac{\partial }{\partial z}\right)+\psi
  \left(-z\frac{ \partial }{\partial z}\right)\right] =\ln \,z+\ln
  \,\frac{\partial }{\partial z}
\]
was used to transform the kinetic part as well as  properties of the
$\psi $-function.

In summary, the exponential mapping eq.(\ref{exp}) 
maps the reggeon dynamics from zero temperature to temperature
$T$. This mapping explicitly exhibits a periodicity 
$\rho \rightarrow \rho +\frac{i}{T}$ for a thermal state. 
It must be noticed that such class
of mappings are known to describe thermal situations for
quantum fields in accelerated frames and in black hole backgrounds\cite{bh}.

\section{Integrability at non-zero temperature}

As it is well known\cite{BKP}, the BFKL equation at $T=0$ can be
generalized to composite states of $n$ reggeized gluons. In
the multi-colour limit $N_{c}\rightarrow \infty $ the BKP equations are
significantly simplified thanks to their conformal invariance,
holomorphic separability and integrals of
motion \cite{lipa}. The generating function for the holomorphic
integrals of motion
coincides with the transfer matrix for an integrable lattice spin model
\cite{integr}. The transfer matrix is the trace of the
monodromy matrix
\[
t(u)=L_{1}(u)\,L_{2}(u)...L_{n}(u)\,,
\]
satisfying the Yang-Baxter equations \cite{integr}. The integrability of the
$n$-reggeon dynamics in multi-colour QCD is valid also at non-zero
temperature $T$, where, according to the above arguments we should
take the $L$-operator in the form
\[
L_{k}=\left(
\begin{array}{cc}
u+p_{k} & e^{-2 \, \pi \, T \, \rho _{k}}\,p_{k} \\
-e^{2 \, \pi \, T \, \rho _{k}}\,p_{k} & u-p_{k}
\end{array}
\right) \,.
\]
In particular, the holomorphic Hamiltonian is the local Hamiltonian of the
integrable Heisenberg model with the spins being unitarily transformed
generators of the M\"{o}bius group (cf. \cite{Heis})
\begin{eqnarray}
&& M_{k}=\partial _{k} \quad ,  \\ \cr
&& M_{+}=e^{-2 \, \pi \, T \, \rho _{k}}\,\partial
_{k} \quad ,  \quad M_{-}=-e^{2 \, \pi \, T \, \rho _{k}}\,\partial _{k}\; .
\nonumber
\end{eqnarray}
Because the Hamiltonian at  non-zero temperature can be obtained by an
unitary transformation from the zero temperature Hamiltonian, the
spectrum of the intercepts
for multi-gluon states is the same as for zero temperature 
and the wave functions of the composite states can be calculated
by the substitution $\rho _{k}\rightarrow \frac1{2 \, \pi \, T} \;
e^{2 \, \pi \, T \; \rho _{k}}$. 

\section{Running with the QCD coupling}
Taking into account the  running of $\alpha_{QCD}$ 
to one-loop level $\alpha_s (Q) = \frac{4 \, \pi}{9 \, \log \frac{4 \, Q^2}{\Lambda^2}} $
changes the previous results. Now, the pomeron wave function must be an eigenfunction of the operator\cite{lip85}
\be\label{ope}
{\cal E} (Q) \equiv \alpha_s (Q) \; \chi\left(- \frac{i}2 
\frac{d}{d \ln |\rho| }\right) \quad , \quad \chi(\nu)= -\frac6{\pi} {\rm Re}\left[\gamma+\psi\left(\frac12+i \, \nu\right)
\right]
\ee
Semiclassically, we have as quantization condition 
\be
\phi(\nu_k) = k + \frac14 \quad , \quad k=0,1,2,\ldots
\ee
where
\be
\phi(\nu) \equiv \frac4{9 \, \alpha_s (Q)}\left[ \frac1{\chi(\nu)} \;
\int_0^{\nu} dx \; \chi(x) - \nu\right] - \frac1{2 \, \pi} \;
\delta_{\nu}^T(\vec{Q}) 
\ee
and
$$
 e^{i \, \delta_{m, {\tilde m}}^T(\vec{Q})} = |Q|^{-4 \, i \, \nu} \;
 \frac{\Gamma(m+i \, Q)}{\Gamma(1-m+i \, Q)}
\frac{\Gamma( {\tilde m}-i \, Q^*)}{\Gamma(1- {\tilde m}-i \, Q^*)}   \; .
$$
The corresponding eigenvalue of the operator $ {\cal E} (Q) $ are given
by
\be
\omega_k(Q) =  \alpha_s(Q) \; \chi(\nu_k)
\ee
The eigenvalues and eigenfunctions now depend parametrically on $ Q 
= \frac{Q_{phys}}{2 \, \pi \, T} $. We recover the zero temperature
limit for $ Q \to \infty $. 
The characteristic scale in temperature is given by $
T_{chara} \equiv \frac1{4 \, \pi} \, e^{\frac{2 \, \pi}{9 \, \alpha_s(Q)}} $
We find that the eigenvalue for $k=0$  first grows
with the temperature and then goes down. In addition, we see that rotational 
invariance is recovered for $ T \to 0 $ since the results for Im Q $= 0 $ and 
$1$ coincide in such limit.

\end{document}